\documentclass{article}

\usepackage{PRIMEarxiv}

\usepackage[utf8]{inputenc} 
\usepackage[T1]{fontenc}    
\usepackage{hyperref}       
\usepackage{url}            
\usepackage{booktabs}       
\usepackage{amsfonts}       
\usepackage{nicefrac}       
\usepackage{microtype}      
\usepackage{lipsum}
\usepackage{fancyhdr}       
\usepackage{graphicx}       
\graphicspath{{media/}}     
\usepackage{amsmath}
\usepackage{caption}
\usepackage{subcaption}
	
\usepackage{soul}
\usepackage{color}
\usepackage{makecell}

\title{Let AI Entertain You: Increasing User Engagement with Generative AI and Rejection Sampling 
}

\author{
  Jingying Zeng, Jaewon Yang, Waleed Malik, Xiao Yan, Richard Huang, Qi He\\
  \vspace{0.01mm}\\
  Nextdoor, USA\\
}

\begin{document}
\maketitle

\begin{abstract}
While generative AI excels in content generation, it does not always increase user engagement. This can be attributed to two main factors. First, generative AI generates content without incorporating explicit or implicit feedback about user interactions. Even if the generated content seems to be more informative or well-written, it does not necessarily lead to an increase in user activities, such as clicks. 
Second, there is a concern with the quality of the content generative AI produces, which often lacks the distinctiveness and authenticity that human-created content possesses.
These two factors can lead to content that fails to meet specific needs and preferences of users, ultimately reducing its potential to be engaging.

This paper presents a generic framework of how to improve user engagement with generative AI by leveraging user feedback. Our solutions employ rejection sampling \cite{nakano2021webgpt}, a technique used in reinforcement learning, to boost engagement metrics. We leveraged the framework in the context of email notification subject lines generation for an online social network, and achieved significant engagement metric lift including +1\% Session and +0.4\% Weekly Active Users.
We believe our work offers a universal framework that enhances user engagement with generative AI, particularly when standard generative AI reaches its limits in terms of enhancing content to be more captivating.
To the best of our knowledge, this represents an early milestone in the industry’s successful use of generative AI to enhance user engagement.
\end{abstract}

\keywords{Generative AI \and Large Language Models \and Social Networks \and Machine Learning \and Reinforcement Learning \and Rejection Sampling}

\section{Introduction}

Generative AI is AI models that create content such as text and images for specific purposes~\cite{bommasani2021opportunities}. Large Language Models (LLMs) are generative AI for textual data. With groundbreaking development in Transformers \cite{vaswani2017attention}, LLMs have achieved remarkable success in downstream tasks such as content summarization, question answering (Q\&A), and content generation, which catalyzed a new era of AI-powered content creation in social media. LLMs such as BERT \cite{devlin2018bert}, GPT-3 \cite{brown2020language}, and LLAMA \cite{touvron2023llama} capture world knowledge through pre-training on massive text data and diverse natural language processing (NLP) tasks \cite{radford2019language, guu2020retrieval}. Riding the wave of generative AI, a vast number of AI-powered applications have been developed in the domains of content creation, assistant application, customer support, conversational agent, and more.


This paper studies how we can increase user engagement in social networks by generating content with LLMs. In other words, we aim to use LLMs to create content that social network users are more likely to click, react, and comment on. With the success of in-context learning \cite{brown2020language}, LLMs have shown promising opportunities in creating different types and styles of content \cite{cao2023comprehensive}. 
To the best of our knowledge, the techniques for generating truly "engaging" content with LLMs have yet to be fully investigated. We know how to use LLMs to produce poetry, for example, but the methods for ensuring that such poetry appeals to a wide audience are still not well-explored.
Increasing user engagement not only invigorates online community interactions but also provides crucial feedback from users. This feedback is pivotal for the continual enhancement and evolution of generative AI models.

Generating engaging content with LLMs is challenging, mainly for the following two reasons.
The first major challenge is that generative AI generates content without incorporating explicit or implicit feedback about user interactions. Even if the generated content seems to be more informative or well-written, it does not necessarily lead to an increase in user activities, such as clicks.  
Although LLMs are trained with certain forms of human feedback~\cite{ouyang2022training}, this is typically aimed at improving adherence to general user-agnostic NLP task instructions, not at increasing the content's interactions with users, which vary widely across social network platforms.

The second major challenge with content produced by generative AI is its quality, which tends to be generic and lacks the distinctiveness and authenticity that human-created content typically possesses.
Recent studies~\cite{guo2023close} discovered that LLMs tend to use a narrower range of vocabulary and adopt a more formal tone than humans typically do. Consequently, the content might be perceived as promotional or similar to advertisements. Additionally, there is the issue of "hallucination", where LLM-generated content with references to non-existent entities or false information, as noted in \cite{manakul2023selfcheckgpt}. To date, there is no known method to entirely eliminate such hallucinations \cite{rawte2023survey}.
Content that is too generic, formal, or factually incorrect can quickly degrade user trust and fails to meet the specific information needs of the user, which can significantly reduce the content's appeal.

\begin{table}[h!]
\centering
\begin{tabular}{|l|l|l|}
 \hline
                             & Rule-generated baseline       & ChatGPT-generated                                     \\
 \hline
Email notification subject line example                & There is a petition in Papillion to & Support backyard chickens  \\
  &  try and get... & in Papillion, NE!\\
  \hline
CTR relative to the baseline & 100\%                                              & 56\% \\                                        \hline
\end{tabular}
\caption{Email notification subject line generated by ChatGPT and its corresponding CTR. GPT's subject line is more informative but looks like a marketing phrase, and produces only 56\% CTR compared to the rule-based subject line, which is simply the first few words of the original user post.}
\label{tab:subject_line_example}
\end{table}

Table~\ref{tab:subject_line_example} shows an illustrative example. In this example, we showcase an email notification featuring a post from Nextdoor, a  leading local social network that brings neighbors and businesses together. We conducted A/B test with two different subject lines: one is simply the first few words of the original user post (Rule-generated baseline in Table \ref{tab:subject_line_example}), and the other is generated by OpenAI's ChatGPT API (ChatGPT-generated in Table \ref{tab:subject_line_example}). Our results showed that subject lines generated by ChatGPT garnered fewer user clicks compared to those generated using simple rules. For the ChatGPT-powered method, we further improved our prompts by giving detailed instructions and adding few-shot examples~\cite{brown2020language}, but the results remained qualitatively similar. This example illustrates that LLMs are not intrinsically designed to generate content that naturally elicits user engagement. 




In this paper, we present a novel framework designed to address the two challenges mentioned earlier. In addition to utilizing an LLM to generate content, we introduce another LLM, the reward model, to assess how much engagement reward the content is likely to receive. We select the content with the highest reward based on our reward model. Through the use of the reward model, we can guide LLMs to compose content that is engaging and interesting to users.

\begin{figure}[h]
    \centering
    \includegraphics[width=0.6\textwidth]{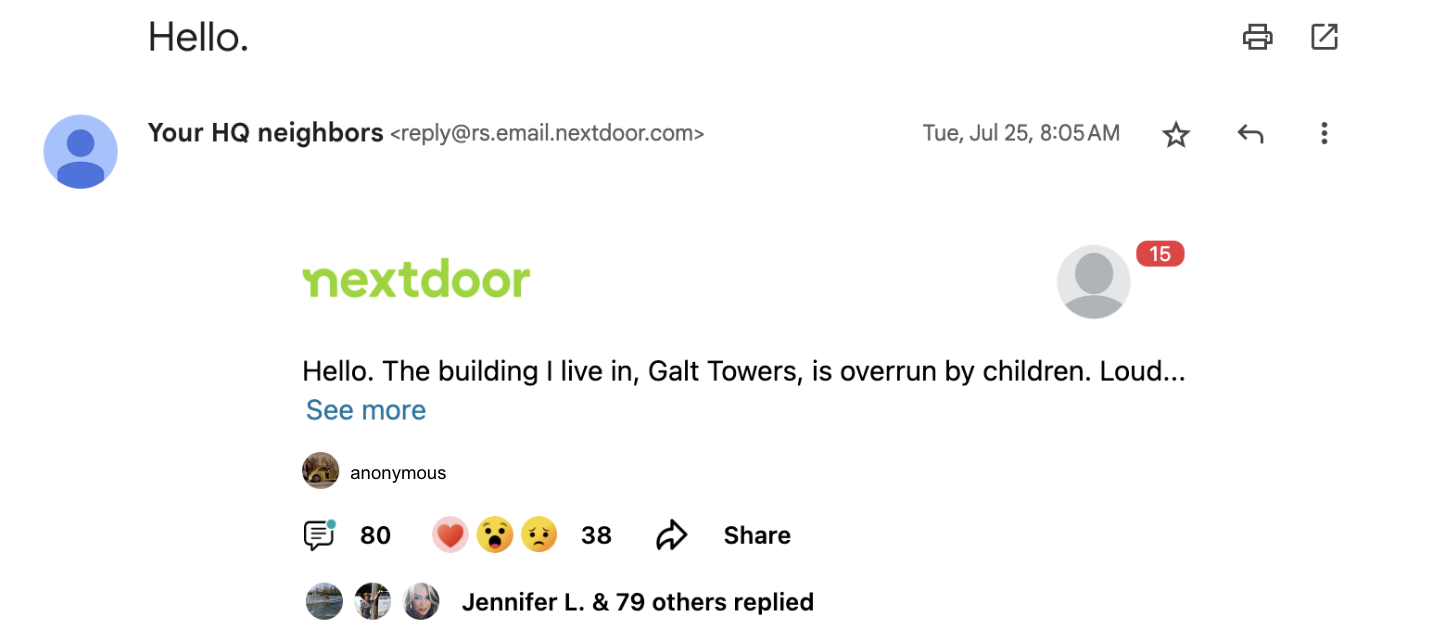}
    \caption{An Example of the Subject Line (Hello.) in a New and Trending Email.}
    \label{fig:control}
\end{figure}

As an application, we tackle the problem of generating subject lines for email notifications on Nextdoor. Nextdoor is the neighborhood network where neighbors, businesses, and public agencies connect with each other. At Nextdoor, we send an email to a user with a single post that the user might be interested in and want to engage with. As part of sending an email, we need to determine a subject line of the email for the email audiences. Historically, we have simply chosen the first few words of the post being sent as the subject line (Figure~\ref{fig:control}). Our goal is to leverage LLMs to compose a more effective subject line.


We developed our methods, deployed to the production on Nextdoor and achieved an engagement lift. Although we showcased our method in one application, we believe our framework can be generally applicable to other social networks. Overall, our contributions are:
\begin{itemize}
    \item \textbf{Implemented a reward model with rejection sampling to explicitly incorporate user feedback}: To steer LLMs toward generating engaging content, we implemented a reward model to predict user preferences for content by using historical user interactions as the training data. The online inference process involves prompting LLMs to produce several variations of text outputs conditioning on the same context, then assessing these variations using the reward model to determine which one is most likely to engage users. We then select the variation of the generated content that is scored the highest by the reward model. This method is known as Best-of-N Sampling, or Rejection Sampling, and has been recently incorporated into the training of LLMs~\cite{nakano2021webgpt, touvron2023llama}. Our analyses highlight the pivotal role of the reward model in improving user engagement. Even with extensive prompt engineering, LLMs without the reward model did not consistently generate more engaging content on their own.
    \item \textbf{Improved content quality by respecting the users' original writing}: Rather than generating entirely new content, we present LLMs with user-generated content and request them to summarize or refine it. In our prompts, we emphasize the importance of preserving the users' original writing. This approach helps eliminate hallucinations and prevent promotional or marketing language in the content.
    \item \textbf{Deployed online in a large social network}: We have deployed an end-to-end engineering system and improved top-line engagement metrics for nearly 85M Nextdoor users. This paper details the methods we used to lower serving costs by caching results, and to ensure steady model performance with daily monitoring, which automatically initiates retraining of the reward model in response to significant shifts in user preferences.
\end{itemize}

\section{User Preference Aware Generator-Evaluator Framework}
Using pre-trained LLM as language generator to generate AIGC that can directly drive user engagement is a challenging task. The quality of a generic pre-trained LLM's generated text can be highly subjective and context dependent. Without explicit training, few-shot in-context learning can adapt the behavior of a generic pre-trained LLM to some extent, but there is still a limit on the amount of possible adaptation. To further improve the quality of generation and better align with human preference, reinforcement learning from human feedback (RLHF) is one of the most common approaches \cite{ouyang2022training}. RHLF first learns a reward model to imitate human judgement and then optimizes a policy model against the reward, using reinforcement learning to improve its performance. While RLHF can effectively align language model (LM) outputs with human feedback, it has not yet been widely accessible to public. On the other hand, collecting labeled user preference data is also challenging since training LMs from human feedback typically requires trained human annotators or human-in-the-loop machine learning platforms for data annotations. 

As an alternative, we propose a generic framework that creates high-quality user feedback from real-time signals, and use it to train both reward and LM-based generator models to encode user preference in content creation. We call this framework a user preference aware generator-evaluator framework. In this framework, either a pre-trained LLM or an LLM supervised fine-tuned using user preference data can be used as an LM-based content generator. Also, a user-preference-encoded reward model is trained to select user preferred texts generated by the LM-based generator model. With user preference embedded into the models, we successfully produce AIGC to drive more engagement metrics when off-the-shelf generative AI models can not be further improved. 

The training framework starts with a pre-trained LLM as our base language generator model. We use in-context learning with few-shot examples to adapt the base generator model to generate task specific textual outputs, and select the best prompt based on empirical evaluation and user feedback through some initial A/B experiments, where the experiments compare the LM-based generator against the rule-based generator in terms of engagement-related metrics such as click-through-rate or thumb-up rate. These live experiments not only aim to select the best prompt for the LM-based generator, but also collect comparison-based preference data for both reward and generator model training. Our training framework uses binary preferences between pairs of examples generated by rule-based and GPT-based generators to learn the latent reward model $r^*_{\theta}$ as a proxy of a human evaluator \cite{zhu2023principled, casper2023open}. Next, we fine-tune the base language generator model, also known as the policy model, and evaluate the performance using the fitted reward model. After the initial round, the fine-tuned generator can be used to generate new training data through new A/B experiments, allowing us to iteratively train the evaluator and generator to continuously optimize user engagements on the LM-generated outputs.

\begin{figure}[h!]
    \centering
    \includegraphics[width=0.9\textwidth]{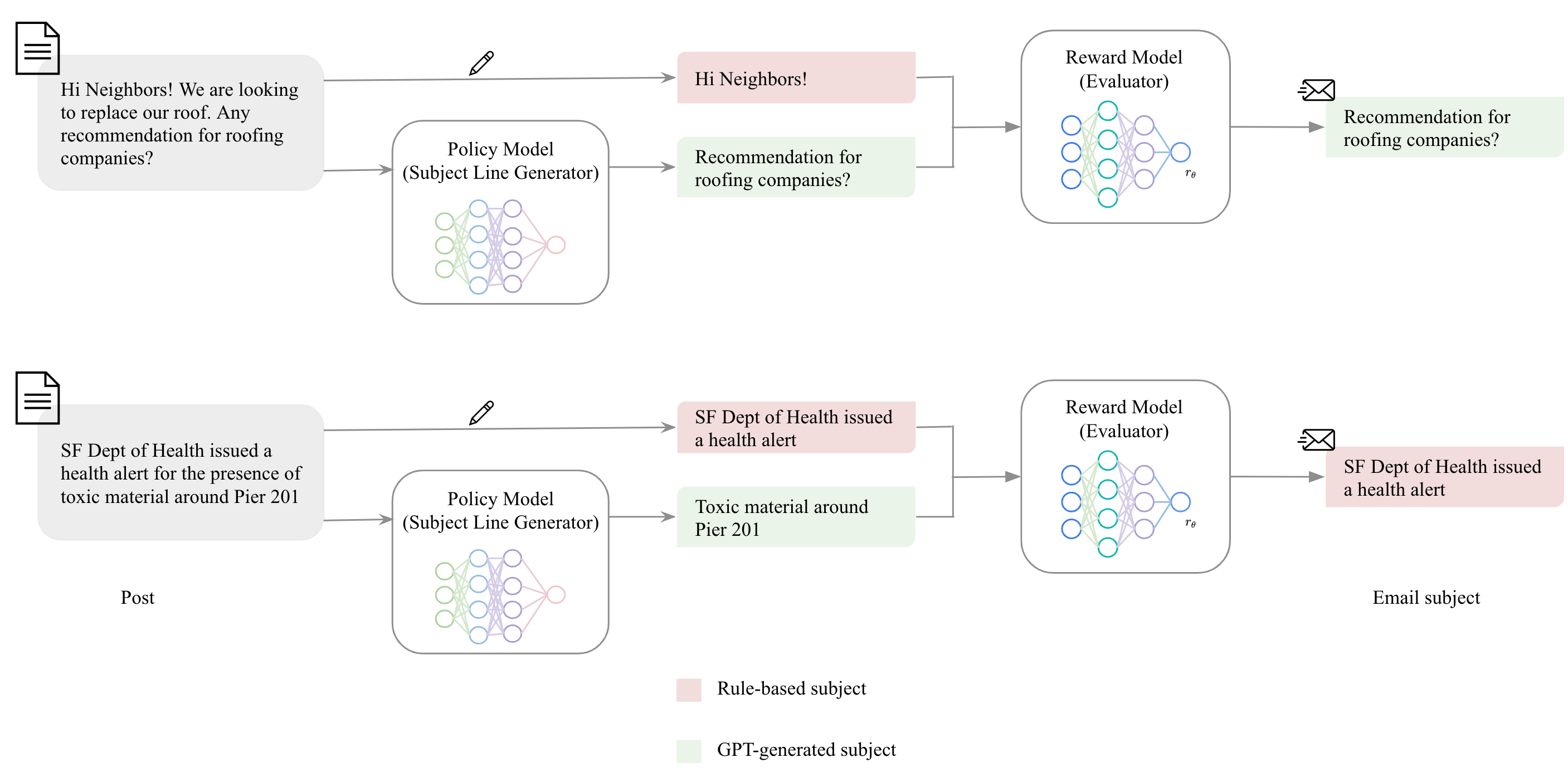}
    \caption{An Overview of the Framework Used in Email Subject Line Generation. Given a post, the subject line generator produces subjects (green boxes). The reward model compares the GPT-generated subject line (green) with the rule-based one (red), and selects the more engaging one between the two. For the example on the top, the GPT-generated subject line is selected by the model since it contains more relevant information. For the example on the bottom, a post about a health alert, the reward model selects the rule-based subject. While the GPT subject line shows the main content of the alert, the reward model picks the rule-based subject because it shows the urgency of the post that tends to be more engaging.}
    \label{reward_policy_diagram}
\end{figure}

For model deployment, we implement caching, retries with fallback, and a daily monitoring system for error handling and cost saving. Figure \ref{reward_policy_diagram} presents a use case of this framework implemented in email subject line generation using a base language generator. The base language generator is a GPT-based generator model. For each post, two subject lines are generated, one from the GPT-based generator and the other from the rule-based generator. The GPT-based generator is prompted by detailed instructions with examples to generate a subject line that is most likely to be appealing to users, while the rule-based generator preserves the user-generated content by extracting the first few words from the user post as an email subject line. With rejection sampling, the more engaging subject line among the two is selected by the reward model and served to users. For the example case shown on the top, the GPT-generated subject line is selected by the reward model since it contains more relevant information than the rule-based one. On the other hand, the example case on the bottom presents an example in which the rule-based subject line is accepted by the reward model since it shows the urgency of the post, which has a higher tendency of being more attractive to users. In the next sections, we will provide more details regarding the training pipeline and model deployment.





\subsection{LM-based Generator Model Optimization}

\textbf{In-context instruction learning for LM generator models:} Without explicit fine-tuning, a popular way to adapt pre-trained LMs to perform new tasks is by "in-context learning" \cite{brown2020language, honovich2022instruction}. Based on a few in-context examples, pre-trained LLMs are able to locate previous learned knowledge to perform new tasks to a certain extent \cite{xie2021explanation, ye2023context}. In the initial round of collecting user feedback from A/B experiments, we use few-shot in-context learning to guide the pre-trained LLM as our language generator to produce desirable outputs that are both authentic and attractive. A series of A/B experiments are conducted to pairwise compare the LM-based and rule-based generated subject lines. The prompt that produces authentic content and boosts the most engagement metrics is selected as the best prompt for the base generator model.

Even though in-context learning is an efficient approach to quickly adapt a pre-trained LM to perform specific tasks, it still reaches a bottleneck in optimizing user engagement. This is expected since pre-training focuses on natural language processing (NLP) tasks such as summarization, instruction following, and etc., instead of improving user engagement. Despite the limitation of in-context learning, pre-trained LLMs can still be used to collect initial user preference data for supervised fine-tuning (SFT) the reward model, use the trained reward model to perform rejection sampling, and present the content generated by LM-based generator only if the reward model approves.

\textbf{Fine-tuning generator via user preference data:} With user preference data, we are able to obtain the user preferred answer for each given context. We then supervise fine-tune the LLM to align the generator model's outputs with user preferences. The fine-tuned generators are offline-evaluated by the trained reward model. The advantage of the fine-tuned language generator compared to the base generator is that the fine-tuned language generator optimizes for user engagement rather than preciseness of summarization. Additionally, the lengths of prompts used in SFT are typically shorter than base generator's, which reduces a significant amount of serving costs.

\subsection{User Preference Aware Reward Model}
A good reward model can be used to approximate a human evaluator in selecting the best-of-N candidates generated by a language model \cite{scheurer2023training} during online serving. Best-of-N is also known as rejection sampling, which selects the sample with the highest reward score among N samples \cite{nakano2021webgpt}. Additionally, a reward model is useful in offline evaluation of the new outputs generated by base models or fine-tuned policy models.

The reward model is trained on user preference data. There are several ways to collect comparison-based user preference data. For instance, LMs can be prompted to produce multiple versions of outputs $\boldsymbol{y} \sim \pi(y | x)$ by conditioning on the prompt $x$. Also, the answers generated by LMs can be compared with the outputs generated by heuristics, such as using the first sentence of the input context. We use the latter method to create binary preference data. After pairs of answers $(y_i, y_j)$ are generated, we then surface them to different treatment groups through live experiments to determine users' implicit preferences via click-through-rates, and create ranked pairs for model training. 

We use SFT to train a reward model that imitates the behavior of a human evaluator. As described below, there are two types of text-to-text models that are commonly used in modeling reward models, namely pointwise reward models and pairwise reward models \cite{scheurer2023training}.

\textbf{Pointwise reward model:} Input sequences of a pointwise reward model combine contexts with generated answers, and the target sequences are either "Yes" or "No". Given context and a generated answer, a pointwise reward model predicts whether or not the answer is a good answer. Mathematically, the reward model is learning the following probability:

\[
P(\text{output sequence} = \text{yes} |x, y_i) \propto P(y_i \succ y_j, \forall{j \neq i} | x)
\]

During inference time, the log probability of outputting the sequence "yes" is used to score and rank generated responses.

\textbf{Pairwise reward model:} In the pairwise approach, user feedback is formulated into a pairwise ranking problem. The input sequences contain the context and a pair of answers, and the target sequences are either "a" or "b".
In other words, given context and answer pair, the pairwise reward model predicts the corresponding winner among them. When formatting the input sequences for the pairwise reward model, the order of the two answers are shuffled randomly. Mathematically, the reward model is learning the following probability \cite{bradley1952rank, rafailov2023direct}:

\[
P(\text{output sequence} = \text{a} |x, y_a, y_b) \propto P(y_a \succ y_b | x) = \frac{\exp(r^*(x, y_a))}{\exp(r^*(x, y_a)) + \exp(r^*(x, y_b))}
\]

At inference time, we use the reward model to rank candidate answers in a list using a tournament style \cite{zhao2023slic}. It calls the reward model $m-1$ times to rank a list of $m$ candidates.

\section{Implementation}
This section describes how we productionized the framework for email subject line generation.
\subsection{Email Subject Line Generation}
We applied the proposed generator-evaluator training framework at Nextdoor to improve user-facing features. One product use case for this framework is email subject line generation. In social networks, a significant portion of users primarily engage with the platform through email notifications. Email notifications are important in keeping users engaged with the community by informing users with relevant and up-to-date posts, replies, mentions, and direct messages. A crucial part of the email notification is the subject line, which provides the "first impression" for users to determine if they want to further engage with the email content or not.

For post-related email notifications, we used two ways to generate email subject lines. One way is using the first sentence of the post with rule-based editing while the other way is using ChatGPT with prompt engineering. We conducted several A/B experiments to compare the rule-based generator against the ChatGPT generator based on the click-through-rate of the linked posts in the emails. Among these pivot studies, we did not notice significant improvements on user engagement using ChatGPT compared to rule-based generator. We found that a more informative subject line generated by ChatGPT does not guarantee a metric lift in clicks. As previously mentioned, a limitation of in-context learning is that prompt engineering alone may not fully capture all the intricacies of the tasks and real users' preferences. To address this, we trained a reward model on the binary preference data collected through previous A/B experiments to predict if the users would prefer a given subject line over other subject lines, and served the best-of-N subject line to the users. We observed a substantial increase in user engagement using this approach. In parallel, we experimented with fine-tuning a pre-trained LM on user preference data and evaluated the generated subject lines offline using the reward model.

\subsection{Model deployment and monitoring}
\textbf{Serving Optimization via caching}: During online serving, in order to mitigate rate limit issues and reduce serving costs, we cache the intermediate and final outputs from the models. Specifically, for a given post, we cache the ChatGPT generated subject lines as intermediate outputs and share cached subject lines among different treatment groups. After the reward model is called, we also cache the winning subject lines for the treatment group that uses the reward model to select the best candidate. By processing each post only once, we reduce the costs by 1/600. In other words, each post is sent 600 times on average, and we process the post only once rather than 600 times. In doing so, we observe a significant drop in the number of tokens used, the number of requests, and the number of rate limit errors from the OpenAI API.


\textbf{Model performance monitoring}: The reward model's predictive performance is monitored daily after the model is deployed. Daily user click data is collected as ground truth and sampled to compare to the reward model's predictions. Specifically, pairwise comparison data is collected from two user buckets, the control bucket and the GPT bucket. Users assigned to the control bucket only receive subject lines generated by the rule-based method, while users assigned to the GPT bucket only receive subject lines generated by ChatGPT. The binary preference data is obtained from these two buckets to measure the reward model's predictive accuracy. If the accuracy drops by more than 10\%, it indicates a shift in user preferences or a shift on the distribution of post content, and we retrain the reward model with the new data.

\textbf{Retries with Fallback}: To further address the rate limit and transient issues from OpenAI API, we add retries with exponential backoffs via Tenacity during online serving. When reaching the cap of the number of retries, we fallback to the rule-based generated subject lines.


\textbf{Controlling the length of output}: The subject line generator sometimes generates subject lines longer than the desired length even when it is instructed to generate with the maximum word limit set to 10. We post-process the generator's output by applying a cut. The optimal maximum length is determined by an A/B experiment on different word limits.


\section{Experiments}

\subsection{Data collection and preprocessing}
User binary preferences are collected from A/B experiments through the relative ranking between two treatment groups. Given a user post, two methods are used in producing email subject lines. One is a rule-based approach, and the other is ChatGPT with a few demonstrations in the prompt. For each post, users are divided into two treatment groups and impressed with one of the two subject lines, as a method to collect user feedback of each method. Side-by-side human evaluation is approximated by the lift ratio of click-through-rates between the two treatment groups. The lift ratio is defined as the click-through-rates of GPT-based generator over rule-based's. We use a predetermined margin $m$ = 0.1 to derive the labels: A lift ratio greater than 1.10 indicates that the GPT-generated subject line is the winner, while a lift ratio less than 0.90 indicates that the rule-based subject line is the winner. 

To mitigate noise in the click data collected from live experiments, we filter posts with $<$ 300 sends and posts with a lift ratio in between 0.90 and 1.10. We also experiment with applying similarity filtering based on the Levenshtein edit distance between two subject lines generated for the same post. However, we find that a high degree of similarity between winner and loser subject lines can be a good signal indicating different user preferences on text features such as emoji or ending style, since the two subject lines might be differed only by an emoji or a "..." at the end. This signal can also provide useful information in training reward model. Therefore, no similarity filtering is applied in the data. The user preference data we collect is then used for training pointwise and pairwise reward models. The final data contains about 50K training examples with 40\% of the examples having GPT subject lines as the winning subject line.

For training the reward model, we format the user preference data into either the pointwise format or the pairwise format using the templates in Table \ref{tab:prompt_reward_model} in the Appendix. For the pairwise format, we randomly shuffle the order of two subject lines for a given post to ensure an equal chance for the rule-based subject line and the GPT-generated subject line to appear first. For training policy model, we format the user preference data into question-answer pairs using the templates shown in Table \ref{tab:prompt_policy_model}.

\subsection{Experimental setups}
\textbf{Experiments on LM-base generator}: The LM-base language generator we use is ChatGPT with in-context instruction learning to guide the LM in generating preferable answers. In the prompt, we instruct ChatGPT to extract the most interesting part from the posts without additional editing, which prevents the generator from hallucinating and generating misleading content. We conduct A/B experiments using different prompts in the base generator and compare its outputs against the rule-based generator's outputs. The pairwise comparison data is collected from these A/B experiments for reward model training.

\textbf{Experiments on reward model}:  We fine-tune different sizes of GPT-3 models and compare the performance of pointwise and pairwise reward models against the baseline gpt-3.5-turbo model (The prompt used for the baseline gpt-3.5-turbo is listed in Table \ref{tab:prompt_baseline_reward_model} in the Appendix). At inference time, we add a logit bias for the tokens "yes" and "no" to increase the probabilities of those tokens appearing. 

In offline analysis, our experiments compare different models based on the accuracy in comparison settings. For the baseline model and the comparison model, the accuracy is computed based on the number of times that the model correctly predicts the winning subject lines. Given a post and a pair of generated subject lines, the pointwise reward model predicts a score for each subject line, defined as the log probability of predicting the token "yes" (i.e. the log probability of predicting a subject line being an engaging subject line). The winning subject line is defined as the one with the higher reward score, and the accuracy is defined by how often the model predicts the winners correctly. We conduct experiments on hyperparameter tuning, increasing the size of the training data, and sensitivity analysis on the thresholds of lift ratio, similarity ratio, and logit bias. 

We also conduct offline evaluation on the performance of best-of-N based on reward scores lifts. Specifically, we experiment with different N and different ways of producing N candidates using the base generator. Two approaches are used when producing N candidates. One is sampling multiple outputs by prompting the base generator conditioning on the same prompt, and the other is using different prompts to generate different outputs.

The final evaluation of the reward model is assessed by A/B experiments based on the metric lifts. We deploy the best-performing reward model with different variations in A/B experiments. The three A/B experiments with different treatment groups we consider are: 
\begin{enumerate}
    \item rule-based subject line v.s. the better subject line between rule-based and ChatGPT, chosen by the reward model
    \item similar to 1 but with different upper limits on the maximum length of subject lines
    \item best-of-N, where $N \in \{1, 3, 5\}$. N different subject line is generated by ChatGPT and the best is chosen by the reward model and served to the users.
\end{enumerate}

\textbf{Experiments on policy model optimization}: The baseline for the fine-tuned policy model is the base generator we use. We evaluate the generated subject lines based on the scores of the best-performing reward model. We calculate the percentage of times that scores from the fine-tuned LM are greater than or equal to the scores from the baseline. We also calculated the percentage of lift in scores, defined as the ratio $r$ of the score average between the fine-tuned LM and the baseline as below:

\begin{equation*}
    r = \mathrm{E}[(p_{fine-tuned})]/\mathrm{E}[(p_{rule-based}].
\end{equation*}

\subsection{Offline and Online Results}
\textbf{Prompt engineering on LM-based generator}: Among the multiple versions of prompt formulation that we experiment through A/B tests, the prompt shown in Table \ref{tab:prompt_subject_line_generator} in the Appendix generates the most engaging subject lines based on the click-through-rates. Not surprisingly, we find that instructing ChatGPT in the prompt to extract subject lines from the content improves sessions by 3.7\% when compared to asking it to generate subject lines from scratch. As expected, prompt engineering improves ChatGPT's performance in generating more user preferable subject lines. However, prompt engineering alone cannot alter the fact that the generator-only method has inferior performance compared to the rule-based one.

\textbf{Accuracy of Reward model:} The accuracy of the baseline reward model is 42.50\%, indicating that predicting user preference is indeed a challenging task. Both pointwise and pairwise reward models are fine-tuned using the smallest GPT-3 model "ada". Table \ref{tab:accuracy_reward_different_epoch} summarizes the accuracy of the pointwise and pairwise reward models trained using epoch \{1, 2, 3, 4\}. We also experiment with fine-tuning on the larger model "Curie", but it does not improve the predictive performance. The pointwise reward model trained using four epochs is the best model, with a predictive accuracy of 66.85\%. We also conduct sensitivity analysis on the best-performing pointwise reward model. We find that the model is not sensitive to the different thresholds of the lift ratio or the similarity ratio. Applying a logit bias of 100 at inference time increases the performance by increasing the chances of the token "yes" or "no" being predicted as the next token. For logit bias, applying different values between 1-100 does not make a significant difference during inference time. However, we observe a significant amount of retry errors when increasing the value beyond 100.

\begin{table}[h!]
\centering
\begin{tabular}{l|llll}
\hline
Epoch                  & 1       & 2       & 3       & 4       \\
\hline
Pointwise reward model & 59.77\% & 61.40\% & 65.06\% & 66.85\% \\
Pairwise reward model  & 61.71\% & 64.98\% & 64.59\% & 64.82\% \\
\hline
\end{tabular}
\caption{Accuracy of Pointwise and Pairwise Reward Model}
\label{tab:accuracy_reward_different_epoch}
\end{table}



\textbf{Analysis of reward model:} When re-formulating the problem as a binary classification of predicting whether the content generated by ChatGPT generator is better and defining the true label based on binary preference feedback, the corresponding precision and recall are 0.67 and 0.37 respectively. This suggests that the model is more likely to pick the rule-based generated subject line over ChatGPT's, which is expected since in the previous A/B experiments show that rule-based methods outperform GPT-based methods, and the reward model is intended to select the better one between the two. On the other hand, among the cases in which the reward model prefers ChatGPT-generated subject line, 67\% of them are also preferred by users. This indicates that reward model is capable of picking high-engaging content from generator. Table~\ref{tab:reward model success cases} below shows some representative cases that the reward model predicts correctly and incorrectly. In the first three examples, the reward model selects the ChatGPT generated subject lines that truly attract more user traffic than the rule-based ones. However, the reward model is not perfect. As shown in the last two examples, when the two subject lines are highly similar to each other, the reward model is not able to correctly predict the more engaging subject lines.

\begin{table}[h!]
  \begin{center}
    \begin{tabular}{ | c | c | c | c | c |}
      \hline
       \thead{id} & \thead{Rule-based Subject Line} & \thead{GPT Subject Line} & \thead{Winner}  & \thead{Model Prediction}\\
      \hline 1 &
Good morning neighbors.                                                                          & \makecell{My boyfriend decided to\\ leave me took the only car...}  &   GPT   & GPT    \\
\hline 2 &
Friends of ours!                                                                                 & \makecell{Condo caught on fire because\\ of a firework landed on...}    &   GPT     & GPT    \\
\hline 3 & \makecell{https://www.kcra.com/article/\\suspected-gunman-deadly-\\roseville-shootout-escapes-custody\\/44484116}  & \makecell{Suspected gunman in deadly \\Roseville shootout escapes custody...}   &   GPT & GPT  \\
      \hline
      4 & \makecell{Mike's Carwash scratched\\ my car} &  \makecell{Mike's Carwash scratched\\ my car...} & GPT & Rule-based\\
      \hline
      5 & \makecell{Please keep your cars\\ locked at all times.} & \makecell{Keep your cars locked\\ at all times...} & Rule-based & GPT\\
      \hline
    \end{tabular}
    \caption{Cases of Reward Model Predictions}
    \label{tab:reward model success cases}
  \end{center}
\end{table}

\begin{figure}[h]
    \centering
    \includegraphics[width=0.7\textwidth]{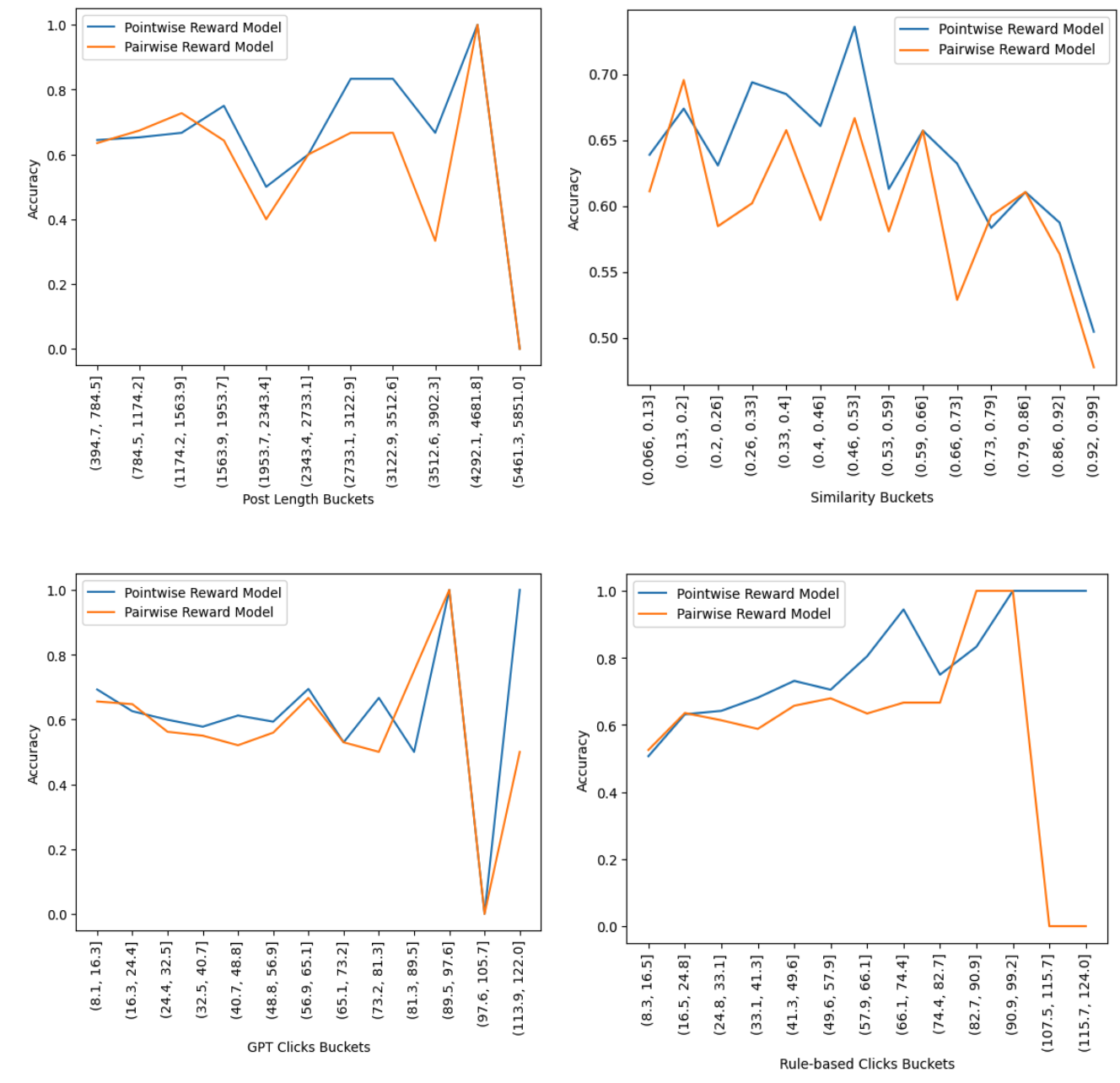}
    \caption{Accuracy Bucketed by Different Dimensions of Similarity, Post Length, Number of Rule-based Clicks, and Number of GPT Clicks.}
    \label{accuracy_diff_dimension}
\end{figure}

Additionally, we compare the accuracy of the pointwise model to the pairwise model on different dimensions such as similarity ratio, length of the posts, and number of clicks. As shown in Figure \ref{accuracy_diff_dimension}, the accuracy of the pointwise model generally outperforms the pairwise model.

\textbf{Analysis of best-of-N:} We first experiment using the same prompt as the template in Table \ref{tab:prompt_subject_line_generator} to generate five versions of email subject lines. Table \ref{tab:best-of-n_same_prompt} summarizes the improvement of best-of-N compared to best-of-1 based on the reward scores. There is an improvement according to reward scores for $N \leq 5$, but after best-of-5, the marginal increase on reward scores is not significant. 

\begin{table}[h!]
\centering
\begin{tabular}{llll}
\hline
best-of-2 & best-of-3 & best-of-4 & best-of-5 \\
\hline
1.18      & 1.29      & 1.36      & 1.42     \\
\hline
\end{tabular}
\caption{Summary statistics of the best-of-N's improvement in reward scores compared to best-of-1 (same prompt)}
\label{tab:best-of-n_same_prompt}
\end{table}

We also evaluate using different prompts to generate multiple versions of subject lines. Table \ref{tab:best-of-n_diff_prompt} summarizes the score improvements.  In addition, we evaluate three versions of prompts based on their reward scores: the prompt as used in the base generator, the prompt without few-shot examples, and the prompt with different few-shot examples. We find that based on the reward model's evaluations, the current prompt achieves the best average scores, which aligns with the A/B experiment results.

\begin{table}[h!]
\centering
\begin{tabular}{llll}
\hline
best-of-2 & best-of-3\\
\hline
1.24      & 1.17 \\
\hline
\end{tabular}
\caption{Summary statistics of the best-of-N's improvements in reward scores compared to best-of-1 (different prompts)}
\label{tab:best-of-n_diff_prompt}
\end{table}

Based on the analysis of best-of-N, including more variants of subject lines in the candidate pool increases the reward model scores. However, one caveat is that without online testing, it is still unknown how much of the offline score increments are correlated with online metrics gain. Therefore, we test it through an online experiment.

\textbf{Finetune policy model}: The policy model is fine-tuned for one epoch using the GPT-3 model "ada". Compared to the rule-based method, the base generator is preferred by reward model in 44.58\% of all test cases,  while the number for fine-tuned generator is 84.29\%. However, we find that 72.11\% out of these 84.29\% generated subject lines are identical to the rule-based ones. Additionally, if we look at the evaluator scores of the subject line from the fine-tuned generator and the base generator, using the fine-tuned generator only increases the score average by 0.4\%. One possible reason is that the data provided to to fine-tune the generator is limited and biased towards rule-based methods (we have more positive examples using rule-based method). Given that no significant performance lift is observed offline with the fine-tuned generator, we decide to use base ChatGPT as the generator in all of our A/B tests. Future data collected from these A/B tests will be used to iterate on the policy model training.


\textbf{A/B experiment results}: To test the framework, we launch A/B experiments with three treatment groups: the rule-based generator, the ChatGPT generator, and the ChatGPT generator with rejection sampling using a reward model as evaluator. As shown in Table \ref{a/b result}, with the generator-evaluator framework, we observe a session lift of 1\%, Weekly Active Users lift of 0.4\%, and Ads revenue lift of 1\%. Additionally, Table \ref{a/b result all model} summarizes the results of all combinations of the generator and reward models. We find that generated subject lines with maximum length of 10 can achieve best performance. However, this length constraint is not observed by the generator if we only use a prompt to enforce it. The A/B experiment of best-of-N has not shown promising results in further driving engagement related metrics compared to best-of-2, which is possibly due to the random noise caused by cache sharing not being implemented for this experiment.

\begin{table}[h!]
\centering
\begin{tabular}{l|lll}
\hline
Metrics   &  Session Lift & Weekly Active Users  & Ads revenue\\
Relative Lift    & 1\%      & 0.4\%                &  1\%\\
\hline
\end{tabular}
\caption{Metrics lift of the final model compared to the rule-based-generated subject lines from A/B tests.}
\label{a/b result}
\end{table}

\begin{table}[h!]
\centering
\begin{tabular}{ll}
\hline
Model      & Session lift \\
\hline
ChatGPT Generator without reward model    &      -6.5\%        \\
ChatGPT Generator (extraction-instructed) without reward model          &      -2.8\%   \\
ChatGPT Generator (extraction-instructed) with reward model         &       +1\%      \\
\hline
\end{tabular}
\caption{Session lift compared to the rule-based-generated subject lines from A/B tests. The final model (last row) achieved 1\% lift in sessions.}
\label{a/b result all model}
\end{table}

\section{Discussion}
Inspired by RLHF, we propose a general framework that utilizes a reward model with rejection sampling to select the best subject lines from the candidate pool and increase user engagement when off-the-shelf generative AI models fail to produce engaging text content. Using a reward model as an evaluator and an LM-generator model with prompt engineering during online serving, we are able to lift engagement-related metrics. 

The keys of making this method successful are as follows. First, when rule-based subject lines only contain introductory remarks without delivering useful information to the users, ChatGPT with proper instructions is able to create more interesting subject lines than a rule-based method is able to. Prompt engineering, on the other hand, enables us to produce more authentic subject lines and decreases the chance of generating marketing phrases or hallucinations. Secondly, we encode user preference into the reward model. Through rejection sampling with a reward model, when GPT-generated subject lines are truly more engaging, the model is able to select the best among a candidate pool. Additionally, the reward model serves as a second guardrail for the outputs generated by ChatGPT to reduce the risk of sending the spam-like subject lines or misleading information that looks like clickbait. In addition to the training framework, serving optimizations also contribute to the success of this framework. Caching the processed posts not only saves costs but also decreases the number of rate limit errors from OpenAI. Additionally, retries with fallback stabilize the error rates from OpenAI.

There are many avenues for future work. First is to fine-tune the subject line generator. In this line of work, we develop the framework using vanilla ChatGPT as the subject line generator in production. We experimented with fine-tuning the generator on binary preference data, but it did not seem to produce more metric lift in our offline analysis. For future work, we can collect k-wise comparison data through A/B experiments by generating multiple subject lines, fine-tune policy model with the most engaging titles that the reward model identifies among the k candidates, and iterate on the fine-tuning process. Secondly, we can incorporating real-time signals into the reward model by re-scoring the same post daily. In the current approach, the reward model can not self-correct since we only use the reward model to pick the best subject line and never re-score. With daily re-scoring, we may be able to see which of the subject lines, GPT-generated or rule-generated, are getting more engagement so as to help the predictive accuracy of the reward model. Thirdly, adding personalization without significantly escalating computational costs is also a research area that we could improve on. The current approach has not taken personalization into account.

\section{Related Work}
\textbf{Large Language Models}: In recent years, decoder-only Large language models (LLMs) such as GPT-4 \cite{openai2023gpt4} and LLaMA \cite{touvron2023llama} have shown great potential in a variety of NLP tasks such as text generation, Q\&A, and reading comprehension. One of the notable capabilities that emerges from the state-of-the-art LLMs is their ability to learn from input-output examples through in-context few-shot learning, without being explicitly trained on specific tasks \cite{brown2020language, xie2021explanation}. In addition to in-context learning, LLMs' instruction-following ability also makes them a powerful tool in personalized content generation. 

Even though LMs achieve remarkable performance on various NLP benchmarks, effectively aligning LMs' outputs with human preferences is still challenging task. The most common way of improving LMs' alignment with human desire is through reinforcement learning from human feedback (RLHF) using high quality annotated human preference data \cite{zhao2023slic}. However, RLHF in general is a complex and expensive process \cite{zhao2023slic, rafailov2023direct} and has not been democratized yet. Given the current off-the-shelf methods, several approaches have been developed as an alternative of RLHF, such as Imitation learning from Language Feedback (ILF) \cite{scheurer2023training}, Direct Preference Optimization (DPO) \cite{rafailov2023direct}. Our framework builds upon these methods developed based on in-context learning and supervised fine-tuning, but uses implicit user engagement data to replace a human annotator in the loop, and combines a reward model as a scorer to perform rejection sampling and simulate human evaluating process during LM content generation.

\textbf{Reward model with rejection sampling}: Supervised fine-tuning a reward model over comparison-based human preference data has been widely used in both academia and industry to reflect human evaluation. A reward model with rejection sampling, also known as best-of-N sampling, is one of the sampling strategies used in reinforcement learning to align model outputs with human desires  \cite{bai2022constitutional, casper2023open}. A series of research has demonstrated the success of using a reward model in real-world applications, including Llama V2 \cite{touvron2023llama}, OpenAI blog, and WebGPT \cite{nakano2021webgpt}, to automate the scoring process.



\textbf{Headline generation}: Headline generation is considered as a branch of text summarization tasks. The literature on headline generation techniques primarily falls into two categories: extractive-based approaches \cite{dorr2003hedge, alfonseca2013heady} and abstractive-based approaches \cite{sun2015event, nallapati2016abstractive, hayashi2018headline}, where extractive-based approaches focus on extracting the important keywords or sentences from the source article, while abstractive-based approaches utilize NLP techniques to summarize the content into one sentence \cite{song2020attractive}. 
With the rise of transformer-based models and the availability of benchmark datasets such as Gigaword \cite{Rush_2015}, CNN Dailymail \cite{see-etal-2017-get}, Newsroom \cite{grusky2018newsroom}, and NewSHead \cite{gu2020generating}, a substantial breakthrough has been achieved in using attention-based models to generate representative headlines \cite{gavrilov2019self, kanungo2022cobart, li2022news}. On the other hand, several studies have investigated personalizing the headline generation by incorporating user profiling \cite{ao2021pens, cai2023generating}, controllable variables (such as length, entity, source-style or document remainders) \cite{fan2017controllable}, and  target
styles \cite{jin2020hooks} to improve the attractiveness of the headlines. Several authors have studied the popularity prediction of headlines \cite{hardt2018predicting} and how to integrate popularity prediction into automatic headline generation \cite{song2020attractive}. While many studies have explored improving the attractiveness of the generated headlines, limited attention has been given to connecting headline generations with boosting engagement-related metrics. Our work bridges this gap by proposing a generic training and deployment framework that utilizes in-context learning to produce authentic and engaging AIGC and uses a reward model trained on comparison-based user preference data to further optimize user engagement.

\section{Conclusion}
In this work, we proposed a user preference aware evaluator-generator framework that can be used to optimize user engagement when in-context learning brings little marginal gains for the LM generator in terms of creating more entertaining content. In this framework, in-context few-shot learning is applied on the base LM-generator to create authentic and user preferred text outputs. A reward model with rejection sampling is used to select the best content generated from different generators. A reward model learns user tastes from binary user preference data collected through live experiments and imitates the auto-evaluation process during serving. LM-generated content will only be accepted if the reward model foresees future user engagement. We test this framework with our serving optimizations and model performance monitoring system on a real-world use case, email subject line generation at Nextdoor. The preliminary results show the success of using AIGC to drive user engagement. This work opens up many possibilities in future research of using generative AI to increase user engagement, ranging from generating more diverse content candidates to personalized LM-generated content.

\section{Acknowledgement}
This work was an collaboration between generative AI and Notification teams at Nextdoor. We extend our appreciation to the following people for their valuable insights and support throughout this project. We would like to thank all the contributors: Hao-Ming Fu, Carolyn Tran, Sameer Suresh, Anna Goncharova, Tiger Zhang.

\bibliographystyle{apalike}  
\bibliography{references}  

\begin{thebibliography}{}

\bibitem[Alfonseca et~al., 2013]{alfonseca2013heady}
Alfonseca, E., Pighin, D., and Garrido, G. (2013).
\newblock Heady: News headline abstraction through event pattern clustering.
\newblock In {\em Proceedings of the 51st Annual Meeting of the Association for Computational Linguistics (Volume 1: Long Papers)}, pages 1243--1253.

\bibitem[Ao et~al., 2021]{ao2021pens}
Ao, X., Wang, X., Luo, L., Qiao, Y., He, Q., and Xie, X. (2021).
\newblock Pens: A dataset and generic framework for personalized news headline generation.
\newblock In {\em Proceedings of the 59th Annual Meeting of the Association for Computational Linguistics and the 11th International Joint Conference on Natural Language Processing (Volume 1: Long Papers)}, pages 82--92.

\bibitem[Bai et~al., 2022]{bai2022constitutional}
Bai, Y., Kadavath, S., Kundu, S., Askell, A., Kernion, J., Jones, A., Chen, A., Goldie, A., Mirhoseini, A., McKinnon, C., et~al. (2022).
\newblock Constitutional ai: Harmlessness from ai feedback.
\newblock {\em arXiv preprint arXiv:2212.08073}.

\bibitem[Bommasani et~al., 2021]{bommasani2021opportunities}
Bommasani, R., Hudson, D.~A., Adeli, E., Altman, R., Arora, S., von Arx, S., Bernstein, M.~S., Bohg, J., Bosselut, A., Brunskill, E., et~al. (2021).
\newblock On the opportunities and risks of foundation models.
\newblock {\em arXiv preprint arXiv:2108.07258}.

\bibitem[Bradley and Terry, 1952]{bradley1952rank}
Bradley, R.~A. and Terry, M.~E. (1952).
\newblock Rank analysis of incomplete block designs: I. the method of paired comparisons.
\newblock {\em Biometrika}, 39(3/4):324--345.

\bibitem[Brown et~al., 2020]{brown2020language}
Brown, T., Mann, B., Ryder, N., Subbiah, M., Kaplan, J.~D., Dhariwal, P., Neelakantan, A., Shyam, P., Sastry, G., Askell, A., et~al. (2020).
\newblock Language models are few-shot learners.
\newblock {\em Advances in neural information processing systems}, 33:1877--1901.

\bibitem[Cai et~al., 2023]{cai2023generating}
Cai, P., Song, K., Cho, S., Wang, H., Wang, X., Yu, H., Liu, F., and Yu, D. (2023).
\newblock Generating user-engaging news headlines.
\newblock In {\em Proceedings of the 61st Annual Meeting of the Association for Computational Linguistics (Volume 1: Long Papers)}, pages 3265--3280.

\bibitem[Cao et~al., 2023]{cao2023comprehensive}
Cao, Y., Li, S., Liu, Y., Yan, Z., Dai, Y., Yu, P.~S., and Sun, L. (2023).
\newblock A comprehensive survey of ai-generated content (aigc): A history of generative ai from gan to chatgpt.
\newblock {\em arXiv preprint arXiv:2303.04226}.

\bibitem[Casper et~al., 2023]{casper2023open}
Casper, S., Davies, X., Shi, C., Gilbert, T.~K., Scheurer, J., Rando, J., Freedman, R., Korbak, T., Lindner, D., Freire, P., et~al. (2023).
\newblock Open problems and fundamental limitations of reinforcement learning from human feedback.
\newblock {\em arXiv preprint arXiv:2307.15217}.

\bibitem[Devlin et~al., 2018]{devlin2018bert}
Devlin, J., Chang, M.-W., Lee, K., and Toutanova, K. (2018).
\newblock Bert: Pre-training of deep bidirectional transformers for language understanding.
\newblock {\em arXiv preprint arXiv:1810.04805}.

\bibitem[Dorr et~al., 2003]{dorr2003hedge}
Dorr, B., Zajic, D., and Schwartz, R. (2003).
\newblock Hedge trimmer: A parse-and-trim approach to headline generation.
\newblock In {\em Proceedings of the HLT-NAACL 03 Text Summarization Workshop}, pages 1--8.

\bibitem[Fan et~al., 2017]{fan2017controllable}
Fan, A., Grangier, D., and Auli, M. (2017).
\newblock Controllable abstractive summarization.
\newblock {\em arXiv preprint arXiv:1711.05217}.

\bibitem[Gavrilov et~al., 2019]{gavrilov2019self}
Gavrilov, D., Kalaidin, P., and Malykh, V. (2019).
\newblock Self-attentive model for headline generation.
\newblock In {\em Advances in Information Retrieval: 41st European Conference on IR Research, ECIR 2019, Cologne, Germany, April 14--18, 2019, Proceedings, Part II 41}, pages 87--93. Springer.

\bibitem[Grusky et~al., 2018]{grusky2018newsroom}
Grusky, M., Naaman, M., and Artzi, Y. (2018).
\newblock Newsroom: A dataset of 1.3 million summaries with diverse extractive strategies.
\newblock {\em arXiv preprint arXiv:1804.11283}.

\bibitem[Gu et~al., 2020]{gu2020generating}
Gu, X., Mao, Y., Han, J., Liu, J., Wu, Y., Yu, C., Finnie, D., Yu, H., Zhai, J., and Zukoski, N. (2020).
\newblock Generating representative headlines for news stories.
\newblock In {\em Proceedings of The Web Conference 2020}, pages 1773--1784.

\bibitem[Guo et~al., 2023]{guo2023close}
Guo, B., Zhang, X., Wang, Z., Jiang, M., Nie, J., Ding, Y., Yue, J., and Wu, Y. (2023).
\newblock How close is chatgpt to human experts? comparison corpus, evaluation, and detection.

\bibitem[Guu et~al., 2020]{guu2020retrieval}
Guu, K., Lee, K., Tung, Z., Pasupat, P., and Chang, M. (2020).
\newblock Retrieval augmented language model pre-training.
\newblock In {\em International conference on machine learning}, pages 3929--3938. PMLR.

\bibitem[Hardt et~al., 2018]{hardt2018predicting}
Hardt, D., Hovy, D., and Lamprinidis, S. (2018).
\newblock Predicting news headline popularity with syntactic and semantic knowledge using multi-task learning.
\newblock In {\em 2018 Conference on Empirical Methods in Natural Language Processing}, pages 659--664. Association for Computational Linguistics.

\bibitem[Hayashi and Yanagimoto, 2018]{hayashi2018headline}
Hayashi, Y. and Yanagimoto, H. (2018).
\newblock Headline generation with recurrent neural network.
\newblock {\em New Trends in E-service and Smart Computing}, pages 81--96.

\bibitem[Honovich et~al., 2022]{honovich2022instruction}
Honovich, O., Shaham, U., Bowman, S.~R., and Levy, O. (2022).
\newblock Instruction induction: From few examples to natural language task descriptions.
\newblock {\em arXiv preprint arXiv:2205.10782}.

\bibitem[Jin et~al., 2020]{jin2020hooks}
Jin, D., Jin, Z., Zhou, J.~T., Orii, L., and Szolovits, P. (2020).
\newblock Hooks in the headline: Learning to generate headlines with controlled styles.
\newblock {\em arXiv preprint arXiv:2004.01980}.

\bibitem[Kanungo et~al., 2022]{kanungo2022cobart}
Kanungo, Y.~S., Das, G., and Negi, S. (2022).
\newblock Cobart: Controlled, optimized, bidirectional and auto-regressive transformer for ad headline generation.
\newblock In {\em Proceedings of the 28th ACM SIGKDD Conference on Knowledge Discovery and Data Mining}, pages 3127--3136.

\bibitem[Li et~al., 2022]{li2022news}
Li, Z., Wu, J., Miao, J., and Yu, X. (2022).
\newblock News headline generation based on improved decoder from transformer.
\newblock {\em Scientific Reports}, 12(1):11648.

\bibitem[Manakul et~al., 2023]{manakul2023selfcheckgpt}
Manakul, P., Liusie, A., and Gales, M.~J. (2023).
\newblock Selfcheckgpt: Zero-resource black-box hallucination detection for generative large language models.
\newblock {\em arXiv preprint arXiv:2303.08896}.

\bibitem[Nakano et~al., 2021]{nakano2021webgpt}
Nakano, R., Hilton, J., Balaji, S., Wu, J., Ouyang, L., Kim, C., Hesse, C., Jain, S., Kosaraju, V., Saunders, W., et~al. (2021).
\newblock Webgpt: Browser-assisted question-answering with human feedback.
\newblock {\em arXiv preprint arXiv:2112.09332}.

\bibitem[Nallapati et~al., 2016]{nallapati2016abstractive}
Nallapati, R., Zhou, B., Gulcehre, C., Xiang, B., et~al. (2016).
\newblock Abstractive text summarization using sequence-to-sequence rnns and beyond.
\newblock {\em arXiv preprint arXiv:1602.06023}.

\bibitem[OpenAI, 2023]{openai2023gpt4}
OpenAI (2023).
\newblock Gpt-4 technical report.

\bibitem[Ouyang et~al., 2022]{ouyang2022training}
Ouyang, L., Wu, J., Jiang, X., Almeida, D., Wainwright, C., Mishkin, P., Zhang, C., Agarwal, S., Slama, K., Ray, A., et~al. (2022).
\newblock Training language models to follow instructions with human feedback.
\newblock {\em Advances in Neural Information Processing Systems}, 35:27730--27744.

\bibitem[Radford et~al., 2019]{radford2019language}
Radford, A., Wu, J., Child, R., Luan, D., Amodei, D., Sutskever, I., et~al. (2019).
\newblock Language models are unsupervised multitask learners.
\newblock {\em OpenAI blog}, 1(8):9.

\bibitem[Rafailov et~al., 2023]{rafailov2023direct}
Rafailov, R., Sharma, A., Mitchell, E., Ermon, S., Manning, C.~D., and Finn, C. (2023).
\newblock Direct preference optimization: Your language model is secretly a reward model.
\newblock {\em arXiv preprint arXiv:2305.18290}.

\bibitem[Rawte et~al., 2023]{rawte2023survey}
Rawte, V., Sheth, A., and Das, A. (2023).
\newblock A survey of hallucination in large foundation models.
\newblock {\em arXiv preprint arXiv:2309.05922}.

\bibitem[Rush et~al., 2015]{Rush_2015}
Rush, A.~M., Chopra, S., and Weston, J. (2015).
\newblock A neural attention model for abstractive sentence summarization.
\newblock {\em Proceedings of the 2015 Conference on Empirical Methods in Natural Language Processing}.

\bibitem[Scheurer et~al., 2023]{scheurer2023training}
Scheurer, J., Campos, J.~A., Korbak, T., Chan, J.~S., Chen, A., Cho, K., and Perez, E. (2023).
\newblock Training language models with language feedback at scale.
\newblock {\em arXiv preprint arXiv:2303.16755}.

\bibitem[See et~al., 2017]{see-etal-2017-get}
See, A., Liu, P.~J., and Manning, C.~D. (2017).
\newblock Get to the point: Summarization with pointer-generator networks.
\newblock In {\em Proceedings of the 55th Annual Meeting of the Association for Computational Linguistics (Volume 1: Long Papers)}, pages 1073--1083, Vancouver, Canada. Association for Computational Linguistics.

\bibitem[Song et~al., 2020]{song2020attractive}
Song, Y.-Z., Shuai, H.-H., Yeh, S.-L., Wu, Y.-L., Ku, L.-W., and Peng, W.-C. (2020).
\newblock Attractive or faithful? popularity-reinforced learning for inspired headline generation.
\newblock In {\em Proceedings of the AAAI Conference on Artificial Intelligence}, volume~34, pages 8910--8917.

\bibitem[Sun et~al., 2015]{sun2015event}
Sun, R., Zhang, Y., Zhang, M., and Ji, D. (2015).
\newblock Event-driven headline generation.
\newblock In {\em Proceedings of the 53rd Annual Meeting of the Association for Computational Linguistics and the 7th International Joint Conference on Natural Language Processing (Volume 1: Long Papers)}, pages 462--472.

\bibitem[Touvron et~al., 2023]{touvron2023llama}
Touvron, H., Martin, L., Stone, K., Albert, P., Almahairi, A., Babaei, Y., Bashlykov, N., Batra, S., Bhargava, P., Bhosale, S., et~al. (2023).
\newblock Llama 2: Open foundation and fine-tuned chat models.
\newblock {\em arXiv preprint arXiv:2307.09288}.

\bibitem[Vaswani et~al., 2017]{vaswani2017attention}
Vaswani, A., Shazeer, N., Parmar, N., Uszkoreit, J., Jones, L., Gomez, A.~N., Kaiser, {\L}., and Polosukhin, I. (2017).
\newblock Attention is all you need.
\newblock {\em Advances in neural information processing systems}, 30.

\bibitem[Xie et~al., 2021]{xie2021explanation}
Xie, S.~M., Raghunathan, A., Liang, P., and Ma, T. (2021).
\newblock An explanation of in-context learning as implicit bayesian inference.
\newblock {\em arXiv preprint arXiv:2111.02080}.

\bibitem[Ye et~al., 2023]{ye2023context}
Ye, S., Hwang, H., Yang, S., Yun, H., Kim, Y., and Seo, M. (2023).
\newblock In-context instruction learning.
\newblock {\em arXiv preprint arXiv:2302.14691}.

\bibitem[Zhao et~al., 2023]{zhao2023slic}
Zhao, Y., Joshi, R., Liu, T., Khalman, M., Saleh, M., and Liu, P.~J. (2023).
\newblock Slic-hf: Sequence likelihood calibration with human feedback.
\newblock {\em arXiv preprint arXiv:2305.10425}.

\bibitem[Zhu et~al., 2023]{zhu2023principled}
Zhu, B., Jiao, J., and Jordan, M.~I. (2023).
\newblock Principled reinforcement learning with human feedback from pairwise or $ k $-wise comparisons.
\newblock {\em arXiv preprint arXiv:2301.11270}.

\end{thebibliography}

\newpage
\section{Appendix}

\begin{table}[h]
\begin{center}
\begin{tabular}{p{0.7\textwidth}p{1\textwidth}}
 \hline \\

 ChatGPT(engine=`gpt-3.5-turbo', temperature=0, max tokens=3, 
 
 \textbf{System message}="You are an assistant to answer which email subject line is more engaging given the Text and subject line a and subject line b. Only answer a or b")

  \textbf{Prompt}: Text: \{post\}
 
  Subject line a: \{subject\_line\_a\}
  
  Subject line b: \{subject\_line\_b\}
  
  Question: Which subject line is more engaging for an email post?
  
  An excellent subject line is coherent, informative, and engaging. We will send this email to our users hoping the users find it interesting and want to click on the email.
  
  Answer with a or b.
  
  Answer:
  
  \#\#\#
 \\
 \hline
\end{tabular}
\end{center}
\caption{Prompt Templates Used in Baseline Reward Model}
    \label{tab:prompt_baseline_reward_model}   
\end{table}

\begin{table}[h]
\begin{center}
\begin{tabular}{p{0.2\textwidth}p{0.6\textwidth}}
 \hline
 Pointwise reward model & Text: \{post\}
 
Subject line: \{subject\_line\}

Question: Is the above an excellent subject line for an email post of the given text?

An excellent subject line is coherent, informative, and engaging. We will send this email to our users hoping the users find it interesting and want to click on the email.

Answer with yes or no.

Answer: 

\#\#\#
 \\ 
 \hline
 Pairwise reward model & Text: \{post\}
 
  Subject line a: \{subject\_line\_a\}
  
  Subject line b: \{subject\_line\_b\}
  
  Question: Which subject line is more engaging for an email post?
  
  An excellent subject line is coherent, informative, and engaging. We will send this email to our users hoping the users find it interesting and want to click on the email.
  
  Answer with a or b.
  
  Answer:
  
  \#\#\#
 \\
 \hline
\end{tabular}
\end{center}
\caption{Prompts Templates Used in Pointwise and Pairwise Reward Model}
    \label{tab:prompt_reward_model}   
\end{table}

\begin{table}[h]
\begin{center}
\begin{tabular}{p{0.7\textwidth}p{1\textwidth}}
 \hline \\
\textbf{Prompt}: We will send an email containing a post from a Nextdoor user. We want to use the most interesting part of the post as an email subject line.

Task description: Given a post, output the most interesting phrase in the post.

Post: \{post\}
 \\
 \hline
\end{tabular}
\end{center}
\caption{Prompt Templates Used in Policy Model}
    \label{tab:prompt_policy_model}   
\end{table}

\begin{table}[h]
\begin{center}
\begin{tabular}{p{1\textwidth}p{1\textwidth}}
 \hline \\
    
 \textbf{Prompt}: We will send an email containing a post from a Nextdoor user. We want to use the most interesting part of the post as an email subject line.

Task description: Given a post, output the most interesting phrase in the post.

Here are the requirements:

1. Extract the phrase as-is. Do not change any single character.

2. Do not paraphrase. Copy the exact phrase. If the phrase you selected has stop words like "but", "and", "the", keep them in the output.

3. Do not insert or remove any word.

3. If you cannot choose the most interesting phrase, return the first 10 words of the post.

5. Try to keep it within 10 words. If you cannot complete within 10 words, generate an incomplete line with "..."

6. Put the most important words in the beginning.

7. If the first 10 words of the post contain unique and interesting words, reuse it.

8. Make a subject line that brings curiosity. If the subject line gets too long, cut the phrase before the last part. For example, if the post has "Yesterday, my son found a dog barking at other people", output "Yesterday, my son found a dog barking at ..."

9. If the first 10 words of the post contain informal words, you can keep these words in the subject line. We want to respect the post content in the subject line. 

10. If the post has a phrase starting with "I" in the first 10 words, please use the same words in the subject line. It will make the subject line more personal. For example, if the post has "Hi All, I left my phone", use "I left my phone" in the subject line.

11. If the some part of the post is all capitals, it is okay to extract that part. That part is what user wanted to emphasize. For example, extract all capital phrases like "CRIME ALERT".

12. Do not use people's names in the subject line.

13. Do not add "Subject line:" in the output. Just output the content of the subject line.

14. Capitalize the first character of the subject line. If the part you selected starts with a lower-cased character, capitalize the character.

These are examples.

Example 1

Post: \{post\}

 \\
 \hline
\end{tabular}
\end{center}
\caption{Prompt Templates Used in Base Subject Line Generator}
    \label{tab:prompt_subject_line_generator}   
\end{table}


\end{document}